\documentclass[prl,aps,twocolumn,showpacs,floatfix]{revtex4}
\usepackage{graphicx}
\usepackage{amsmath}
\begin{document}
\title{Anomalous Hall Effect in Ferromagnetic Semiconductors in the Hopping 
Transport Regime}
\author{A.A. Burkov and Leon Balents}
\affiliation{Department of Physics, University of California, Santa Barbara, 
California 93106, USA}
\date{\today}

\begin{abstract}
We present a theory of the Anomalous Hall Effect (AHE) in ferromagnetic
(Ga,Mn)As in the regime when conduction is due to phonon-assisted 
hopping of holes between localized states in the impurity band. 
We show that the microscopic origin of the anomalous Hall conductivity in 
this system can be attributed to a phase that a hole gains when hopping 
around closed-loop paths in the presence of spin-orbit interactions and 
background magnetization of the localized Mn moments.
Mapping the problem to a random resistor network, 
we derive an analytic expression for the macroscopic 
anomalous Hall conductivity $\sigma_{xy}^{AH}$. 
We show that $\sigma_{xy}^{AH}$ is proportional to the first derivative
of the density of states $\varrho(\epsilon)$ and thus can be expected 
to change sign as a function of impurity band filling.
We also show that $\sigma_{xy}^{AH}$ depends on temperature as the 
longitudinal conductivity $\sigma_{xx}$ within logarithmic accuracy.  

\end{abstract}

\maketitle

Diluted magnetic semiconductors (DMS) are very interesting new materials,
still not understood in full detail. 
Although it is well accepted that 
the origin of ferromagnetism in Ga$_{1-x}$Mn$_x$As, the most thoroughly 
studied DMS, is a hole-mediated effective
interaction between the localized $S=5/2$ Mn spins, the appropriate 
description of this interaction, and a closely related question of the 
nature of hole carriers, are still controversial issues.
Samples with the highest Curie temperatures show metallic 
behavior \cite{Potashnik01} and are well described
by the valence-band hole-fluid model \cite{holefluid}.
On the other hand, for low Mn concentrations, i.e. when $x$ is less
than about $0.03$, 
(Ga,Mn)As is insulating and the hole transport likely occurs in 
the Mn impurity band \cite{impurityband}.
The low-temperature conduction in this case is by phonon-assisted hopping.

Anomalous Hall Effect (AHE) has been an important characterization tool 
for itinerant ferromagnets \cite{Smit,Luttinger58,Berger70}
and played an important role in the experimental study of DMS ferromagnetism
as well \cite{Ohno}.
Experimentally AHE is manifested as an additional term in the Hall 
resistivity of the sample, usually assumed to be proportional to the 
magnetization:
\begin{equation}
\label{eq:1}
\varrho_{xy}=R_0 B + R_s M.
\end{equation}   
Standard theories of AHE in metallic ferromagnets \cite{Smit,Berger70}
attribute it to the modification of impurity scattering in the presence 
of spin-orbit interactions. 
Recently, however, it was realized that a purely geometrical 
mechanism of AHE is 
possible \cite{MacDonald02,Millis99,Lyanda,Nagaosa}.
In particular, a ${\bf k}$-space Berry phase theory of AHE \cite{MacDonald02} 
has been proposed, and applied specifically to (Ga,Mn)As in the 
metallic transport regime. 
This theory is very successful in describing AHE in metallic (Ga,Mn)As,
and we believe that the Berry-phase mechanism of AHE should be relevant 
in general for all itinerant ferromagnets.

In this Letter we propose an analogous geometrical-phase theory of AHE in 
(Ga,Mn)As, but in the case when conduction is by hopping 
between strongly localized states in the impurity band, rather than through
extended valence-band states. 
Our approach is based on a modified Holstein's theory of the Hall effect 
in hopping conduction \cite{Holstein61} (for related work on AHE in colossal 
magnetoresistance manganites see \cite{Lyanda}).

We start from a model of spin-3/2 hole carriers, hopping between
randomly distributed localizing centers (Mn acceptors) in the mean 
spin-splitting field of $S=5/2$ Mn local moments.
The corresponding Hamiltonian is:
\begin{equation}
\label{eq:2}
H_h = \sum_{i\alpha} \epsilon_i c^{\dag}_{i\alpha} 
c_{i\alpha} - 
\sum_{i\alpha j\beta}t_{i\alpha j\beta} c^{\dag}_{i\alpha} c_{j\beta}
-\sum_{i\alpha\beta} {\bf h} \cdot \boldsymbol{\tau}_{\alpha\beta} 
c^{\dag}_{i\alpha} c_{i\beta}. 
\end{equation}     
Here $i,j$ label localized hole states at Mn acceptor sites,
$\alpha,\beta=3/2,\ldots,-3/2$ are spin-3/2 indices,
$\epsilon_i$ are random on-site potentials, which model the random 
potentials localized holes feel due to Coulomb interactions with nearby 
charged Mn acceptors and As antisite defects, ${\bf h}$ is the spin-splitting
mean field due the p-d exchange interaction between the hole- and localized 
Mn-spins, $\boldsymbol{\tau}$ is the spin-3/2 operator and 
$t_{i\alpha j\beta}$ is a spin-dependent hopping amplitude which
we assume to be of the following form:
\begin{equation}
\label{eq:3}
t_{i\alpha j\beta}={\cal R}_{ij} \,\hat t\, 
{\cal R}^{\dag}_{ij}\,e^{iA_{ij}-r_{ij}/\xi},
\end{equation}
where $\xi$ is the localization length and
$A_{ij}=(e/2 c){\bf B} \cdot ({\bf r}_i \times {\bf r}_j)$ is a phase 
factor due to the external magnetic field ${\bf B}$ (we use $\hbar=1$ units).
In most of the following we will omit this factor.
$\hat t=diag(t_{3/2},t_{1/2},t_{1/2},t_{3/2})$ is a diagonal 
hopping matrix in the case when $\hat r_{ij}$ is along the 
spin quantization direction and ${\cal R}_{ij}$ rotates the spin quantization
axis from $\hat z$ to $\hat r_{ij}$.
The hopping parameters $t_{3/2}$ and $t_{1/2}$ can be calculated 
microscopically \cite{Fiete02}. The fact that $t_{3/2} \neq t_{1/2}$ is a
direct consequence of the spin-orbit interactions.  

We assume that the Fermi energy lies in the region where all states are 
strongly localized and thus the only possible conduction mechanism is through
the phonon-assisted hopping. 
Thus hole-phonon interaction must be explicitly included in the model:
\begin{eqnarray}
\label{eq:4}
&&H_{h-ph}=\sum_{i\alpha{\bf q}} v^i_{\bf q} (b_{\bf q}+
b^{\dag}_{-{\bf q}})
c^{\dag}_{i\alpha}c_{i\alpha},\nonumber \\
&&H_{ph}=\sum_{\bf q}\omega_{\bf q} b^{\dag}_{\bf q} b_{\bf q}.
\end{eqnarray}
The full Hamiltonian is then given by
\begin{equation}
\label{eq:5} 
H=H_h+H_{h-ph}+H_{ph}.
\end{equation}

Local transport properties of the system described
by Eq.(\ref{eq:5}) can be most conveniently calculated using a combination 
of the usual linear-response theory and a perturbation expansion in powers of 
$t_{i\alpha j\beta}/|\epsilon_i-\epsilon_j|$, which is justified by our 
assumption that all states are 
localized \cite{Entin95,Galperin96}.
Holstein \cite{Holstein61} was first to realize that to capture the 
(ordinary) Hall effect under hopping conduction conditions, one needs to 
consider processes where amplitudes of direct and indirect (through a third
site) hops between two sites interfere.
Such processes appear at third order in the perturbation theory in
$t_{i\alpha j\beta}/|\epsilon_i-\epsilon_j|$, while the longitudinal 
hopping conductivity can be described by second-order terms.
To capture AHE, one needs to consider third-order terms as well.
Evaluating the linear response expression for the current $I_{ij}$ between 
two localized states \cite{Galperin96}, including terms of up to third order 
in the hopping amplitudes and up to second order in hole-phonon interaction, 
one obtains:
\begin{equation}
\label{eq:6}
I_{ij}=G_{ij}V_{ij}+\sum_k F_{ijk}(V_{ik}+V_{jk}).
\end{equation}
Here $V_{ij}=V_i-V_j$ is the electrochemical potential difference between
the sites $i$ and $j$. 
$G_{ij}$ is an effective conductance between a pair of sites,
which at low temperatures, such that $|\epsilon_i-\epsilon_j|/T \gg 1$, is 
given by:
\begin{equation}  
\label{eq:7}
G_{ij}=\frac{e^2 \gamma_0}{T} \sum_{\alpha\beta} |t_{i\alpha j\beta}|^2
e^{-\epsilon_{i\alpha j\beta}/T}.
\end{equation}
Here $\gamma_0$ is a constant, proportional to the phonon density of states
and the square of the hole-phonon interaction matrix element, and we have 
introduced the following notation:
\begin{eqnarray}
\label{eq:8}
&&\epsilon_{i\alpha j\beta} = \frac{1}{2}\left(|\epsilon_{i\alpha}| + 
|\epsilon_{j\beta}| + |\epsilon_{i\alpha}-\epsilon_{j\beta}|\right), 
\nonumber \\
&& \epsilon_{i\alpha}=\epsilon_i - h \tau^z_{\alpha \alpha}.
\end{eqnarray} 
We have assumed in the above that ${\bf h}= h \hat z$. 
Henceforth we will also assume that $h$ is sufficiently small, so that we 
can restrict ourselves to {\it linear response} effects with respect to $h$.
In particular this means that we disregard the influence of spin splitting on 
the longitudinal conduction. 
We expect this to be a reasonable approximation at low Mn concentrations, 
when the magnetization is relatively small. 

The first term in Eq.(\ref{eq:6}) thus represents the usual mapping of the 
hopping conduction problem onto a random resistor 
network \cite{Miller60,Shklovskii71,Langer71}.
The second term in Eq.(\ref{eq:6}) contains the physics 
responsible for AHE.
The quantity $F_{ijk}$ is given, in the low temperature 
limit, by the following expression:
\begin{eqnarray}
\label{eq:9}
&&F_{ijk}=\frac{e^2 \gamma_0^2}{8 T} \sum_{\alpha\beta\gamma} 
\Im(t_{i\alpha j\beta}t_{j\beta k\gamma}t_{k\gamma i\alpha})
\left[e^{(|\epsilon_{k\gamma}|-\epsilon_{i\alpha k\gamma}-
\epsilon_{k\gamma j\beta})/T}\right. \nonumber \\ 
&&\left.+e^{(|\epsilon_{i\alpha}|-\epsilon_{k\gamma i\alpha}-
\epsilon_{i\alpha j\beta})/T}+
e^{(|\epsilon_{j\beta}|-\epsilon_{i\alpha j\beta}-
\epsilon_{j\beta k\gamma})/T}\right].
\end{eqnarray}
Let us discuss the physical meaning of this equation.
The process, that leads to the appearance of the term, proportional 
to $F_{ijk}$, in the expectation value of the current between sites $i$ 
and $j$, is a quantum interference process, where probability amplitudes 
of a direct 
hop between any two sites in a triad $(i,j,k)$ and an indirect hop 
(i.e. through the third site) interfere.
Such a process requires the participation of at least two phonons, hence the 
factor $\gamma_0^2$. 
The crucial component of $F_{ijk}$ is the imaginary part of the product 
of three complex spin-dependent hopping amplitudes taken along the sides 
of the triangle formed by the sites $i,j,k$.
It is directly related to the phase that a hole gains while hopping 
along a closed-loop three-site path. 
It is this phase that ultimately leads to the appearance of the macroscopic
anomalous Hall conductivity. 
Note that if the mean spin splitting field $h$ is taken to zero, 
$\sum_{\alpha\beta\gamma} 
\Im(t_{i\alpha j\beta}t_{j\beta k\gamma}t_{k\gamma i\alpha})$ vanishes 
identically for any triad $(i,j,k)$. 
It is the presence of a macroscopic magnetization in the sample,
or, in other words, broken time reversal symmetry, that results 
in a nonzero contribution from this term.
Another important property of the third-order term in the current 
expectation value is that it leads to the appearance of a current between the 
sites $i$ and $j$, that is driven not by the potential difference on the bond 
$(ij)$, but by the potential differences on bonds $(ik)$ and $(jk)$.
This property implies that the current, proportional to $F_{ijk}$, is 
on average transverse to the direction of the local electric field, as 
one would expect for the Hall current.

Eq.(\ref{eq:6}) reveals the microscopic origin of the AHE in the system. 
To obtain the anomalous Hall conductivity of a macroscopic sample, 
one needs to find a certain characteristic microscopic triad that 
determines the macroscopic conductivity.
Since the distributions of both $G_{ij}$ and $F_{ijk}$ are very broad due to 
their exponential dependence on the random on-site potentials and Mn acceptor
positions, the characteristic triad can not be obtained by a simple averaging 
over all possible triads.
Instead, one needs to find a relationship between 
this generalized hopping conduction problem and a certain percolation
problem, exactly like in the case of usual hopping 
conduction \cite{Shklovskii71,Langer71}.
It is, however, not an easy task, due to a tensor nature of the transport
brought about by the second term in Eq.(\ref{eq:6}).
We believe that the best approach to this problem is the one  
used in Refs. \cite{Bryksin77,Pollak,Galperin} to investigate the 
problem of ordinary Hall effect in hopping conduction regime.
This approach starts from the assumption that the problem can be solved
perturbatively in the second, Hall term in Eq.(\ref{eq:6}). 
This is certainly a valid assumption if one wants to find the  
Hall conductivity to first order in the magnetic field (or magnetization
in our case).   
Thus one imagines the following two-step iterative procedure.
Initially, the second term in (\ref{eq:6}) is put to zero and external 
electric field is applied to the network.
One then solves Kirchhoff's equations
\begin{equation}
\label{eq:10}
\sum_j G_{ij}V^0_{ij}=0
\end{equation} 
and finds voltages $V^0_i$ at all nodes.
The external electric field is then turned off and the zeroth-order voltages
are substituted in the Hall term of Eq.(\ref{eq:6}).
The Kirchhoff's equations at the second iteration take the following form:
\begin{equation}
\label{eq:11}
\sum_j G_{ij}V_{ij}=J^H_i,
\end{equation}
where
\begin{equation}
\label{eq:12}
J^H_i = - \sum_{jk} F_{ijk}(V^0_{ik}+V^0_{jk}),
\end{equation}
i.e. now we have a network with external Hall 
currents $J^H_i$ flowing into every node.
Thus the paradigm of the hopping Hall conduction in the approach we adopt
is a {\it random resistor 
network with external Hall currents} \cite{Bryksin77}.
 
\begin{figure}
\includegraphics[width=4cm]{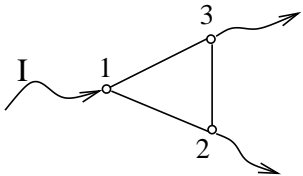}
\caption{A backbone node with an associated triad. 
$I$ is the current flowing in a macrobond terminating at site 1 in response 
to the external electric field $E$.} 
\label{triad}
\end{figure}
To actually obtain explicit analytical results using the above procedure, 
one needs to know the distribution of Hall currents $J^H_i$ in the network. 
We will use ideas from the percolation theory picture of hopping 
transport~\cite{Shklovskii71,Langer71,Shklovskii74} to find this distribution.
According to this picture, one can assume that all the hopping current 
in the network is carried by a percolating cluster of large conductances, such
that for all conductances in the cluster $G_{ij} \geq G_c$, where 
$\ln G_c \sim -(T_0/T)^{1/4}$ (we will limit our discussion to Mott's
variable-range-hopping type conduction, but our theory, with minor 
modifications, is applicable
to other kinds of hopping transport as well). 
The longitudinal conductivity of the network is then given by
$\sigma_{xx} \approx G_c/L$, where $L$ is the correlation length of the
cluster.
The correlation length depends on temperature only as a power 
law~\cite{Shklovskii74} and we will not calculate it explicitly. 
Following Refs. \cite{Bryksin77,Pollak,Galperin}, we now assert that the Hall 
conductivity in the 
system is associated exclusively with the current carrying backbone 
of the percolating cluster. 
The backbone can be visualized as a supernetwork of {\it macrobonds} and
{\it nodes}, with typically three macrobonds intersecting at a node.
The average distance between the nodes is $L$.
Macrobonds can be assumed to be simply one-dimensional chains of 
conductances \cite{Shklovskii74}.
It is then clear that the exponentially dominant Hall current sources
will be located at the nodes of the percolating cluster backbone. 
Since the nodes are far away from each other,
we may regard them as independent sources of the Hall emf.
Thus the problem reduces to finding the Hall emf developed at a single
isolated node of the backbone in response to the current flowing 
through the node \cite{Bryksin77,Pollak,Galperin}.
Applying the above described two-step iterative procedure to a single
node with an associated triad of sites shown in Fig.\ref{triad}, we obtain:
\begin{eqnarray}
\label{eq:14}
V^H \equiv V_{32} &=&\frac{G_{13} J^H_2 - G_{12} J^H_3}
{G_{13}G_{32} + G_{12}G_{23} + G_{31}G_{12}}\nonumber\\ 
&=&\frac{I F_{123}}{G_{13}G_{32} + G_{12}G_{23} +
G_{31}G_{12}},
\end{eqnarray}
where $V^{H}$ is the Hall emf, $I$ is the current flowing through the node, 
and we have neglected unimportant numerical prefactors.
Since the average distance between the nodes is $L$, the macroscopic 
Hall field in the sample is given by:
\begin{equation}
\label{eq:15}
E^H = I L^{-1} \left\langle \frac{S_{123} F_{123}}
{G_{13}G_{32} + G_{12}G_{23} + G_{31}G_{12}} \right\rangle_c,
\end{equation}
where $\langle \ldots \rangle_c$ denotes averaging over the on-site energies 
and intersite distances in the triad $(1,2,3)$, such that the condition
\begin{equation}
\label{eq:16}
2 r_{ij}/\xi + \epsilon_{ij}/T \leq 
\left(T_0/T\right)^{1/4}
\end{equation}
is satisfied for every bond $(ij)$ in the triad~\cite{Langer71} and
$S_{123}=\textrm{sgn} [{\bf h}\cdot({\bf r}_{12}\times {\bf r}_{13})]$ 
accounts for the 
orientation of the triangle with respect to ${\bf h}$.
Taking into account that the Hall current density is given by 
$j^H = \sigma_{xx} E^H$ and $I = \sigma_{xx} L^2 E$, we obtain the following 
expression for the anomalous Hall conductivity \cite{Bryksin77,Pollak}:
\begin{equation}
\label{eq:17}
\sigma_{xy}^{AH} = L \sigma_{xx}^2 \left\langle \frac{S_{123} F_{123}}
{G_{13}G_{32} + G_{12}G_{23} + G_{31}G_{12}} \right\rangle_c.
\end{equation}
Let us write down the averaging explicitly as an integral over 
intersite distances and on-site energies:
\begin{eqnarray}
\label{eq:18}
&&\sigma_{xy}^{AH} = L \sigma_{xx}^2 \frac{1}{\cal N}
\int_c d{\bf r}_{12} d{\bf r}_{23} 
d{\bf r}_{31}\int_c d \epsilon_1 d\epsilon_2 d \epsilon_3\nonumber\\ 
&&\times\varrho(\epsilon_1) \varrho(\epsilon_2) \varrho(\epsilon_3)
\frac{S_{123} F_{123}}{G_{13}G_{32} + G_{12}G_{23} + G_{31}G_{12}},  
\end{eqnarray}
where ${\cal N}$ is a normalization factor.
The densities of states $\varrho(\epsilon)$ can be expanded around the 
Fermi energy 
$\varrho(\epsilon) \approx \varrho_0+\frac{d\varrho_0}{d\epsilon}\epsilon$, 
since hopping is restricted
to a narrow interval of order $T(T_0/T)^{1/4}$ around the Fermi energy.
Substituting the expanded densities of states in Eq.(\ref{eq:18}), one finds 
that the zeroth order term, proportional to $\varrho_0^3$, vanishes.
The reason is that $F_{123}$ changes sign under the inversion of all the 
on-site energies, while the bond conductances $G_{ij}$ are invariant 
under this transformation.
Thus we obtain:
\begin{equation}
\label{eq:19}
\sigma_{xy}^{AH} = L \sigma_{xx}^2 \frac{d \ln \varrho_0}{d\epsilon} 
\left\langle \frac{S_{123} F_{123} (\epsilon_1 + \epsilon_2 + \epsilon_3)}
{G_{13}G_{32} + G_{12}G_{23} + G_{31}G_{12}}\right\rangle_c,  
\end{equation}
where the average is now over a uniform distribution of the on-site energies.
Approximating the average of the quantity inside the angular brackets by its 
maximum value and expanding to 
first order in $h$, one finally 
obtains the following expression for the anomalous Hall hopping conductivity:
\begin{equation}
\label{eq:20}
\sigma_{xy}^{AH} \sim L \sigma_{xx}^2 \frac{d \ln \varrho_0}{d\epsilon} 
\frac{h \, T}{e^2 t_{3/2}} 
\left(T_0/T\right)^{1/4}\,e^{-(T_0/T)^{1/4}},
\end{equation}     
where we have used the property that $t_{3/2} \gg t_{1/2}$ in GaMnAs 
\cite{Fiete02}.
Analogous calculation in the case of the ordinary Hall effect gives
\begin{equation}
\label{eq:21}
\sigma_{xy}^{OH} \sim L \sigma_{xx}^2 \frac{B\,\xi^2\, T}{c\, e\, t_{3/2}}\,
e^{-(T_0/T)^{1/4}}.
\end{equation}
Thus, the ratio of the ordinary and anomalous Hall conductivities is
\begin{equation}
\label{eq:22}
\frac{\sigma_{xy}^{AH}}{\sigma_{xy}^{OH}} \sim h\,
\frac{d \ln \varrho_0}{d\epsilon} \frac{c}{e B \xi^2}
\left(T_0/T\right)^{1/4}.
\end{equation}
Several concrete experimentally testable predictions follow from the above 
equations.
First, according to (\ref{eq:20}), $\sigma_{xy}^{AH}$ is proportional to the 
derivative of the density of states at the Fermi energy. 
This means that the anomalous Hall conductivity can be expected to 
change sign as the Fermi level crosses the density-of-states maximum
in the impurity band.
This is in contrast to the ordinary Hall conductivity $\sigma_{xy}^{OH}$ 
which has the same sign everywhere in the impurity band. 
Thus AHE can potentially be a very useful tool in elucidating 
the character of hole carriers in (Ga,Mn)As.
The second prediction concerns the temperature dependence of 
$\sigma_{xy}^{AH}$.
As seen from Eq.(\ref{eq:20}), the leading temperature dependence 
of the anomalous Hall conductivity is 
$\sigma_{xy}^{AH} \propto \exp[-(T_0/T)^{1/4}]$, i.e. it depends on 
temperature in the same manner as the longitudinal conductivity $\sigma_{xx}$.
The same is true for the ordinary Hall conductivity.
This, in particular, means that the Hall resistivity 
$\varrho_{xy} \sim \varrho_{xx}$ as a function of temperature, i.e.
it diverges as $T \to 0$. 
Thus our calculation is consistent with the view \cite{Entin95} that the 
transverse resistivity of the ``Hall insulator'' diverges at $T \to 0$.

In conclusion we would like to point out that our theory seems to be consistent
with recent magnetotransport experiments on digitally-doped GaMnAs samples
\cite{Gwinn}.   

We are grateful for useful discussions with W. Allen, E. Gwinn, G. Fiete
and E. Johnston-Halperin.
We would especially like to thank D.D. Awschalom for his encouragement and 
support.
This work was supported by DARPA/ONR N00014-99-1-1096, by the NSF under 
grant DMR-9985255 and by the Sloan and Packard foundations.

\end{document}